\documentclass[twocolumn,10pt]{article}

\usepackage[T1]{fontenc}
\usepackage[utf8]{inputenc}
\usepackage{lmodern}
\usepackage{graphicx}
\usepackage{amsmath,amssymb}
\usepackage{braket}
\usepackage[super,sort&compress]{natbib}
\usepackage[a4paper,margin=1.8cm]{geometry}
\usepackage{xcolor}

\title{An Additive Reference Correction Scheme for the Transcorrelated Method}
\author{Johannes Hauskrecht, Kristoffer Simula, Yifan Cheng, Evelin Martine Corvid Christlmaier,\\
Daniel Kats, and Ali Alavi\\[0.5em]
\small Max Planck Institute for Solid State Research, Heisenbergstra\ss e 1, 70569 Stuttgart, Germany\\
\small Yusuf Hamied Department of Chemistry, University of Cambridge, Lensfield Road, Cambridge CB2 1EW, UK \\
\small \texttt{j.hauskrecht@fkf.mpg.de}, \texttt{k.simula@fkf.mpg.de}, \texttt{y.cheng@fkf.mpg.de},\\
\small \texttt{e.christlmaier@fkf.mpg.de}, \texttt{d.kats@fkf.mpg.de}, \texttt{a.alavi@fkf.mpg.de}}
\date{}

\begin{document}

\twocolumn[
\maketitle
\begin{abstract}
We introduce an additive reference correction for the transcorrelated (TC) method and its three-body mean-field approximation (xTC), to improve energy differences computed in small orbital basis sets. The correction is motivated by the observation that, for xTC atomization energies, the dominant error in double-$\zeta$ bases originates from the reference contribution rather than from the correlation energy. In the proposed reference-corrected scheme (RC-xTC), the small-basis correlation energy is retained, while the corresponding TC reference energy is replaced by its value from a larger basis. Benchmark calculations for the non-relativistic HEAT set with the Dunning basis-set family show that RC-xTC substantially improves both total and atomization energies relative to standard xTC in double-$\zeta$ bases. At the CCSD(T) level, RC-xTC yields better atomization energies than CCSD(T)-F12a  in the double-$\zeta$ regime, while preserving the favorable total-energy accuracy of xTC. At the CCSD level, RC-xTC improves atomization energies relative to F12a throughout the full basis-set sequence. As the basis set is enlarged, xTC and RC-xTC become progressively identical, as expected from the construction of the correction.
\end{abstract}
\vspace{0.5cm}
]

%\section{Introduction}

The slow basis-set convergence of conventional orbital expansions in electronic-structure theory originates largely from their poor description of the short-range electron--electron cusp. Explicitly correlated approaches address this difficulty by incorporating interelectronic distances more directly. Among them, the transcorrelated (TC) method, originally introduced by Boys and Handy, rewrites the Schr\"odinger equation using a similarity transformation with a Jastrow factor, thereby transferring a substantial part of the short-range correlation problem from the wave function into an effective Hamiltonian.\cite{boys1969neon,handy1969be}

In recent years, TC methods have undergone a marked revival. Recent work has combined TC Hamiltonians with stochastic and deterministic many-body solvers, including FCIQMC for the homogeneous electron gas and molecular applications,\cite{luo2018tcfciqmc,haupt2023jastrow,christlmaier2023xtc,filip2025secondrow} transcorrelated DMRG for ground and excited molecular states,\cite{baiardi2020density,liao2023dmrg} and transcorrelated coupled-cluster formulations, including recent extensions to non-iterative triples corrections.\cite{schraivogel2021tcc,schraivogel2023tcc2,kats2024orbital} TC ideas have also been developed in periodic and solid-state contexts, in renewed formal studies of bi-variational TC wave functions, in biorthonormal orbital optimization for quantum Monte Carlo and transcorrelation, and in compactification strategies for determinant expansions.\cite{ochi2023tcpp,tsuneyuki2008tcsolids,simula2026tcper,luo2026tdl,lee2023studies,ammar2021biorthogonal,ammar2024compactification} These developments have established TC and xTC approaches as promising explicitly correlated routes to high-accuracy total energies in comparatively modest basis sets.

For thermochemical applications, the quality of the error cancellation between related species is often more important than the absolute accuracy of the individual total energies. In our experience, and as will be documented below, this balance can remain insufficiently robust in double-$\zeta$-quality orbital bases, even when TC/xTC total energies are already quite accurate. This issue becomes especially apparent when comparing against the F12 framework. Explicitly correlated F12 methods routinely recover near-complete-basis-set quality with compact orbital bases and have proved successful in thermochemical protocols such as W1-F12, W2-F12, and W4-F12.\cite{kong2012f12,karton2012w1f12,sylvetsky2016w4f12} An important ingredient in that context is the treatment of one-particle incompleteness in the reference contribution. In practical F12 implementations, this is commonly addressed through the complementary auxiliary basis set (CABS) singles correction,\cite{shaw2017cabs} while closely related ideas already appear in early CCSD(T)-F12 developments\cite{adler2007ccsdtf12} and in analyses of one-particle basis-set relaxation in R12/F12 theories.\cite{noga2009oneparticle}

Motivated by this idea, we introduce an additive reference correction for the transcorrelated method. The aim is not to modify the small-basis TC correlation treatment itself, but rather to correct the reference contribution in a way that improves the cancellation of basis-set errors between atoms and molecules. We assess the resulting scheme on the HEAT benchmark set,\cite{tajti2004heat} which provides a stringent test for atomization energies.

%The remainder of this paper is organized as follows. Section~2 derives and motivates the reference correction scheme. Section~3 presents benchmarking results for the HEAT set. Section~4 summarizes the main findings and discusses implications for future TC applications.

%\section{Motivation and reference-correction scheme}

\begin{figure}[!t]
  \centering
  \includegraphics[width=\linewidth]{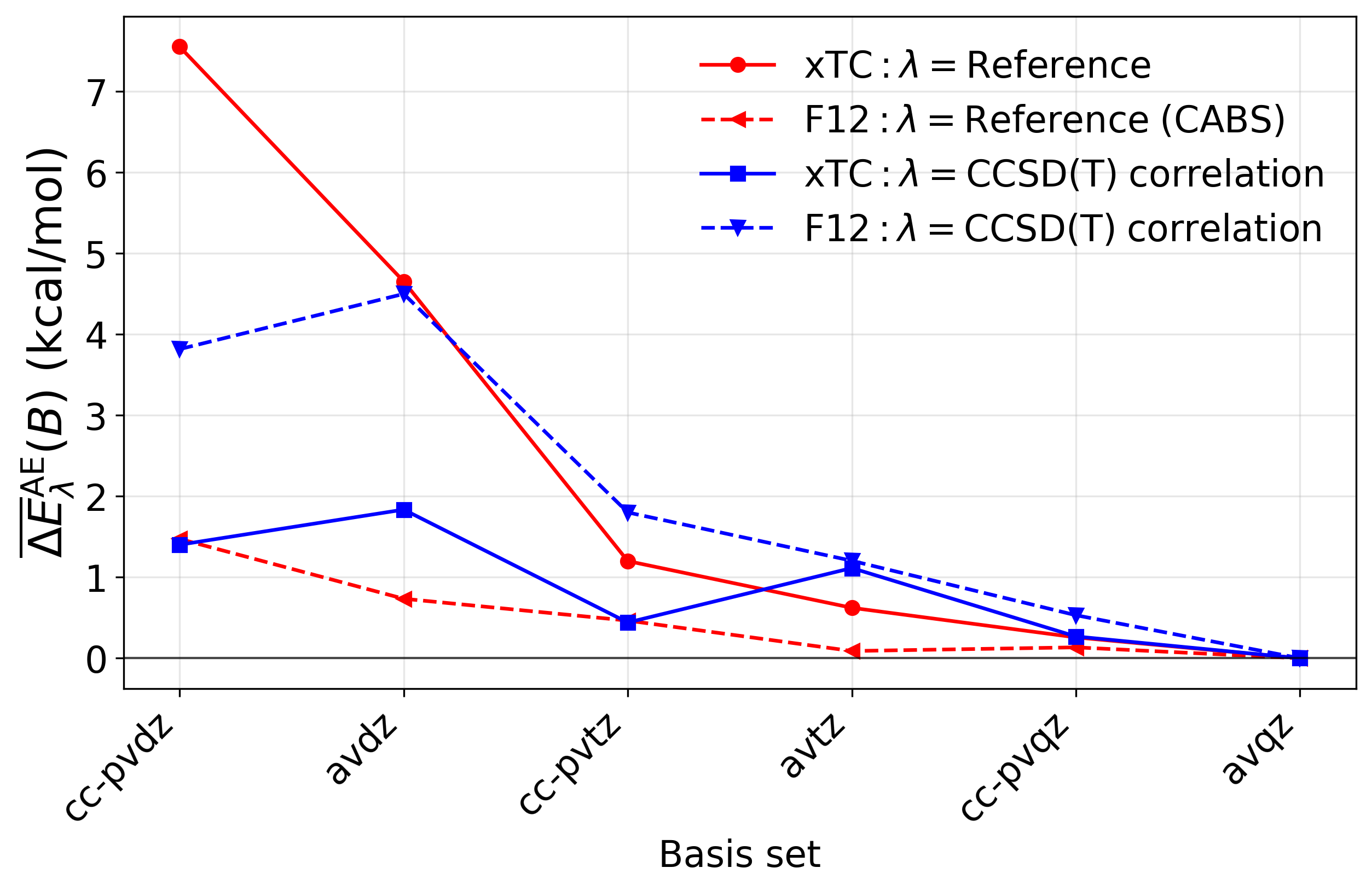}
\caption{
Basis set dependent mean absolute errors for the HEAT set of the atomization-energy deviations defined in Eqs.~(\ref{eq:Delta_E_ato}) and (\ref{eq:Mean_Delta_E_ato}), relative to aug-cc-pVQZ. Shown are the reference and correlation contributions across the Dunning basis-set family. The underlying energies are obtained with xTC-CCSD(T) (solid lines) and CCSD(T)-F12 (dashed lines). For the xTC data, the Jastrow factor was optimized in aug-cc-pVQZ and then reused throughout the corresponding basis-set family.
}

  \label{fig:mae_total}
\end{figure}

We first analyze the origin of the atomization-energy errors in xTC. Our primary target in this work is the improvement of relative energies. While xTC already yields very accurate total energies in modest orbital bases, the error cancellation required for relative energies is more demanding and, as shown in Fig. \ref{fig:mae_total} for atomization energies, is limited mainly by the reference contribution in small basis sets.

The TC Hamiltonian is given by
\begin{equation}
\bar{H}_{\mathrm{TC}}(\tau) = e^{-\tau} H e^{\tau} = H + [H,\tau] + \frac{1}{2}[[H,\tau],\tau],
\end{equation}
where $\tau$ is a real function, which in this study is of Drummond--Towler--Needs (DTN) form,\cite{drummond2004jastrow,haupt2023jastrow}
\begin{equation}
\tau = \tau_{ee} + \tau_{en} + \tau_{een},
\end{equation}
with polynomial channels and cutoff functions for the electron--electron, electron--nucleus, and electron--electron--nucleus terms.

In practice, we will employ two approximations to make the TC scheme practical. First, we employ the xTC approximation, in which the three-body terms which arise from the double-commutator are contracted down to two-body form, 
\begin{equation}
\bar{H}_{\mathrm{xTC}}(\tau) = H + [H,\tau] + \frac{1}{2}[[H,\tau],\tau]_{12},
\end{equation}
the equations of which are provided by Christlmaier et al \cite{christlmaier2023xtc}. This approximation has been shown to introduce only very minor error with respect to the full TC Hamiltonian \cite{liao2021periodic,schraivogel2021tcc, schraivogel2023tcc2}, whilst drastically simplifying the formalism. Second, in order to solve the xTC Hamiltonian, we will most often employ the CCSD(T) method, adapted to the non-hermitian nature of the xTC Hamiltonian,\cite{hino2001biorthogonal,kats2024orbital} referring to the corresponding energies $E^{\mathrm{xTC}}$ as the energy of xTC-CCSD(T). Of course, this method is, by its nature, an approximation to the true ground-state energy of the xTC Hamiltonian. A broader aim of transcorrelation, however, is to improve the intrinsic accuracy of such approximations relative to the corresponding non-transcorrelated (i.e.\ standard) CC methods, as has been suggested previously.\cite{schraivogel2021tcc,schraivogel2023tcc2} We will examine this point further in light of the proposed improvements of the present work.

To define the quantities discussed below, we first introduce the total xTC energy in basis $B$, associated with the correlated xTC wave function $\Psi_B$, as
\begin{equation}
E^{\mathrm{xTC}}(\tau;B)
=
\langle \Psi_B | \bar{H}_{\mathrm{xTC}}(\tau) | \Psi_B \rangle
.
\label{eq:xtc_total}
\end{equation}
We then decompose this energy as
\begin{equation}
E^{\mathrm{xTC}}(\tau;B)
=
E_{\mathrm{ref}}^{\mathrm{xTC}}(\tau;B)
+
E_{\mathrm{corr}}^{\mathrm{xTC}}(\tau;B),
\label{eq:xtc_decomp}
\end{equation}
with
\begin{equation}
E_{\mathrm{ref}}^{\mathrm{xTC}}(\tau;B)
=
\langle \Phi_0(B) | \bar{H}_{\mathrm{xTC}}(\tau) | \Phi_0(B) \rangle,
\label{eq:xtc_ref}
\end{equation}
where $\Phi_0(B)$ denotes the reference determinant in basis $B$ and $E_{\mathrm{corr}}^{\mathrm{xTC}}(\tau;B)$ is the corresponding correlation energy. The decomposition itself is general and does not depend on the specific post-Hartree--Fock method used to approximate the correlated xTC wave function $\Psi_B$.
The Jastrow parameters are determined stochastically by minimizing the variance of the TC reference energy,\cite{haupt2023jastrow}
\begin{equation}
\sigma_{\mathrm{ref}}^2(\tau;B)
=
\langle \Phi_0(B) |
\left|
\bar{H}_{\mathrm{TC}}(\tau)
-
E_{\mathrm{ref}}^{\mathrm{TC}}(\tau;B)
\right|^2
| \Phi_0(B) \rangle.
\end{equation}
For comparison, on the F12 side we write
\begin{equation}
E^{\mathrm{F12}}(B)
=
E_{\mathrm{ref}}^{\mathrm{F12}}(B)
+
E_{\mathrm{corr}}^{\mathrm{F12}}(B),
\end{equation}
where the reference contribution is taken as the Hartree--Fock energy plus the complementary auxiliary basis set (CABS) singles correction, \cite{shaw2017cabs, adler2007ccsdtf12, noga2009oneparticle}
\begin{equation}
E_{\mathrm{ref}}^{\mathrm{F12}}(B)
=
E_{\mathrm{HF}}(B)+\delta E_{\mathrm{CABS}}(B),
\end{equation}
with
\begin{equation}
\delta E_{\mathrm{CABS}}(B)
=
\sum_{i\alpha}\frac{|F_{i\alpha}|^2}{\varepsilon_i-\varepsilon_\alpha}
=
\sum_{i\alpha} F_{i\alpha}\, t_i^\alpha,
\label{eq:cabs}
\end{equation}
and $E_{\mathrm{corr}}^{\mathrm{F12}}(B)$ is the corresponding correlation contribution.

For each basis $B$ and each energy component $\lambda\in\{\mathrm{ref},\mathrm{corr}\}$, we define the componentwise atomization energy of a molecule $M$ as
\begin{equation}
E^{\mathrm{AE}}_{\lambda}(M;B)
=
\sum_{A\in M} n_A E_{\lambda}(A;B) - E_{\lambda}(M;B),
\end{equation}
where $n_A$ denotes the stoichiometric multiplicity of atom $A$. The corresponding deviation from the aug-cc-pVQZ reference is
\begin{equation}
\Delta E^{\mathrm{AE}}_{\lambda}(M;B)
=
E^{\mathrm{AE}}_{\lambda}(M;B)-E^{\mathrm{AE}}_{\lambda}(M;B_{\mathrm{QZ}}),
\label{eq:Delta_E_ato}
\end{equation}
with $B_{\mathrm{QZ}}=\mathrm{aug\text{-}cc\text{-}pVQZ}$. Figure~\ref{fig:mae_total} reports the mean absolute errors (MAEs) for the HEAT set of the deviations defined in Eq.~(\ref{eq:Delta_E_ato}),
\begin{equation}
\overline{\Delta E}^{\mathrm{AE}}_{\lambda}(B)
=
\frac{1}{N}
\sum_{M\in\mathrm{HEAT}}
\left|
\Delta E^{\mathrm{AE}}_{\lambda}(M;B)
\right|,
\label{eq:Mean_Delta_E_ato}
\end{equation}
using energies obtained with xTC-CCSD(T) and CCSD(T)-F12.\cite{tajti2004heat}

\begin{figure*}[!t]
 \centering
    \begin{minipage}{0.49\textwidth}
    \centering
    \includegraphics[width=\linewidth]{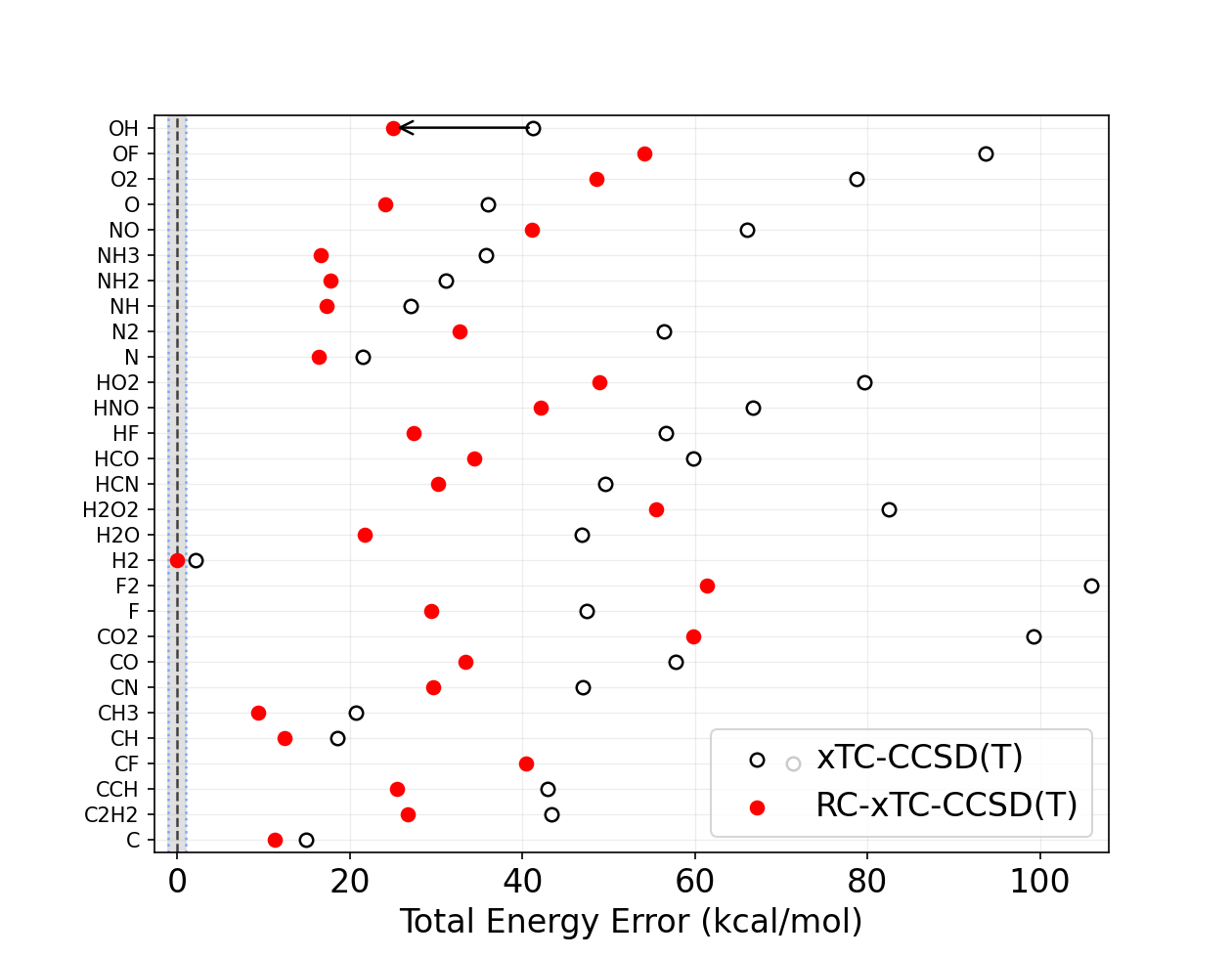}
  \end{minipage}\hfill
  \begin{minipage}{0.49\textwidth}
    \centering
    \includegraphics[width=\linewidth]{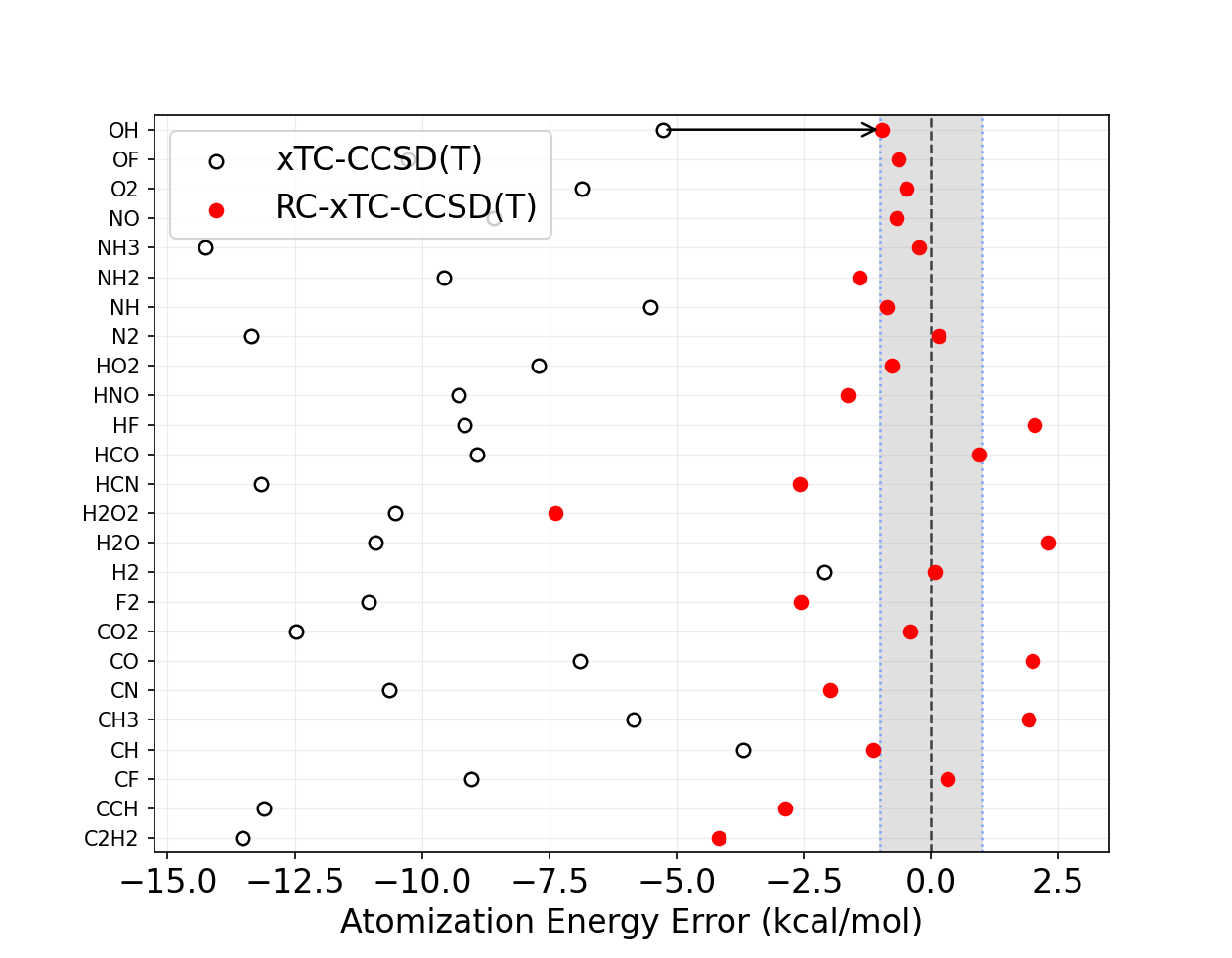}
  \end{minipage}
  \caption{
  System-resolved total-energy (left) and atomization-energy errors (right) for xTC-CCSD(T) and RC-xTC-CCSD(T) in the cc-pVDZ basis. For every HEAT system, the reference-corrected scheme lowers the total-energy error. The same improvement carries over to the atomization energies, demonstrating that the improved total energies lead directly to better error cancellation. The MAEs and MaxEs for this data can be found in Table \ref{tab:dunning_stats}.
  }
  \label{fig:ccsdt_scatter}
\end{figure*}

The need for an xTC correction analogous in spirit to the CABS correction of Eq.~(\ref{eq:cabs}) is illustrated in Fig.~\ref{fig:mae_total}. The figure reports the MAEs of the reference and correlation contributions to the atomization energies across the Dunning basis-set family, relative to aug-cc-pVQZ. For the xTC data, the Jastrow factor was optimized with Hartree--Fock reference wave functions in aug-cc-pVQZ and then reused for the smaller bases.

This decomposition is highly revealing. In the double-$\zeta$ regime, {\em the dominant xTC error in atomization energies stems from the reference contribution rather than from the correlation contribution}. Indeed, the xTC correlation error is already smaller than the corresponding CCSD(T)-F12 correlation error, whereas on the F12 side the CABS-corrected reference error is smaller than the correlation error. Thus, the dominant deficiency of xTC atomization energies in small basis sets is not the correlation treatment itself, but the reference contribution. This strongly suggests that further improvement should focus on correcting the xTC reference energy.

\begin{figure*}[!t]
  \centering
  \includegraphics[width=\textwidth]{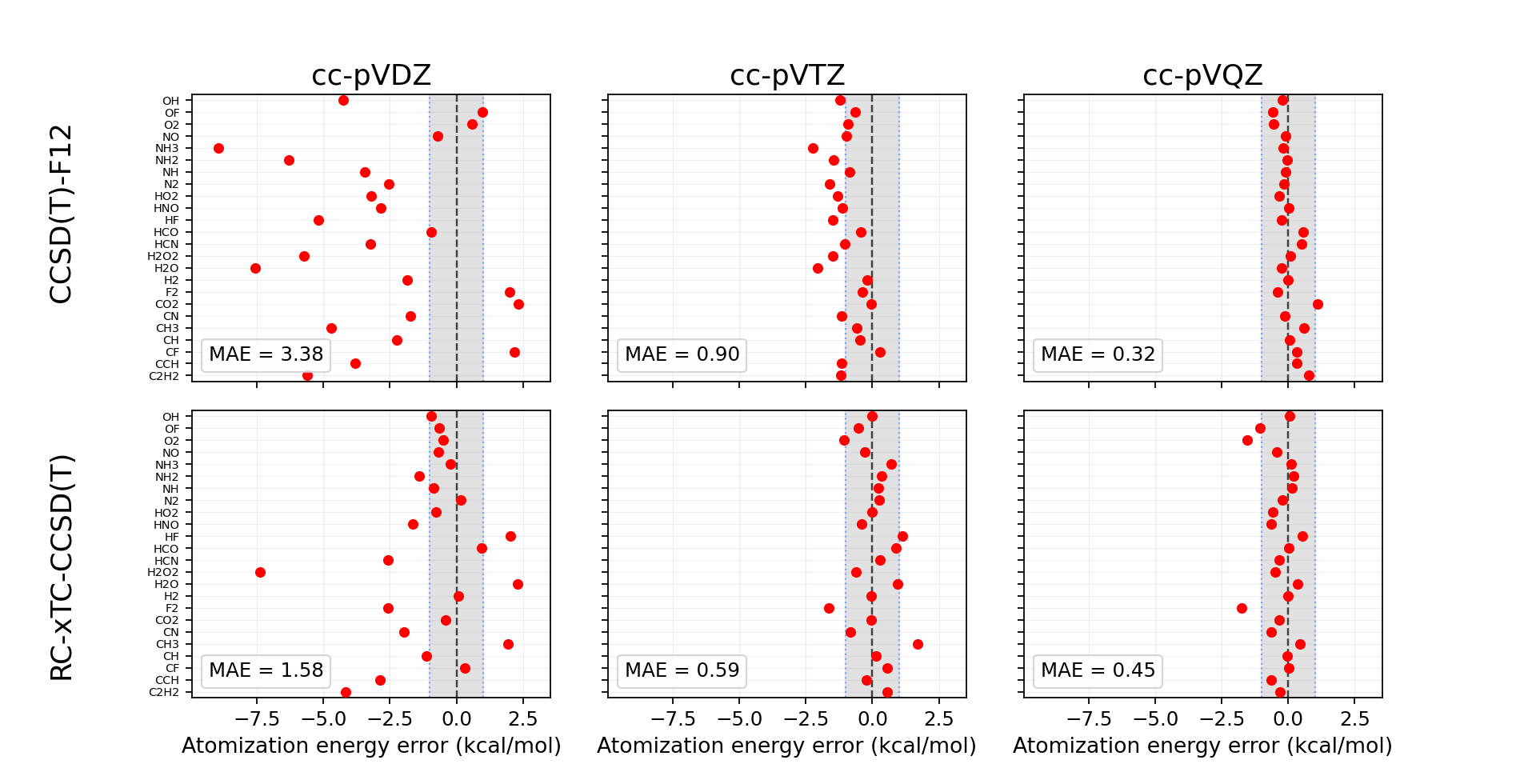}
\caption{
Atomization-energy errors for the HEAT systems at the CCSD(T) level, comparing CCSD(T)-F12 (top row) and RC-xTC-CCSD(T) (bottom row) for cc-pVDZ, cc-pVTZ, and cc-pVQZ. In cc-pVDZ, the spread of errors is smaller for RC-xTC than for F12. With increasing basis size, both methods become more compact, but in cc-pVTZ a larger fraction of systems is already chemically accurate for RC-xTC, whereas CCSD(T)-F12 generally requires cc-pVQZ to reach a comparable number of chemically accurate results.
}
 \label{fig:ccsdt_basis}
\end{figure*}

Motivated by this observation, we introduce a simple additive reference correction. Let $S$ and $L$ denote a small and a larger basis, respectively, and let $\tau_L$ be a Jastrow factor optimized in the larger basis $L$. We do not require $S$ to be a strict subspace of $L$; rather, $L$ is chosen such that its reference energy is substantially closer to the complete-basis-set limit than that of $S$, ideally essentially converged with respect to one-particle basis-set effects. Using the decomposition in Eq.~(\ref{eq:xtc_decomp}), we define the reference shift 

\begin{align}
\Delta E_{\mathrm{ref}}^{\mathrm{xTC}}(\tau_L;S,L)
&=
E_{\mathrm{ref}}^{\mathrm{xTC}}(\tau_L;L)
-
E_{\mathrm{ref}}^{\mathrm{xTC}}(\tau_L;S).
\label{eq:refshift}
\end{align}
The reference corrected small-basis energy is then obtained by replacing the small-basis reference contribution with its large-basis counterpart evaluated for the same Jastrow:
\begin{align}
E^{\mathrm{RC\text{-}xTC}}(\tau_L;S,L)
&=
E^{\mathrm{xTC}}(\tau_L;S)
+
\Delta E_{\mathrm{ref}}^{\mathrm{xTC}}(\tau_L;S,L)
\nonumber\\
&=
E_{\mathrm{ref}}^{\mathrm{xTC}}(\tau_L;L)
+
E_{\mathrm{corr}}^{\mathrm{xTC}}(\tau_L;S).
\label{eq:refcorr}
\end{align}

In other words, the correlation energy is retained from the small basis $S$, while the reference energy is imported from the larger basis $L$. Motivated by the decomposition analysis above, this construction directly targets the dominant source of atomization-energy error seen in Fig.~\ref{fig:mae_total}. Thus, both the CABS correction and the present RC-xTC scheme act, in spirit, as additive improvements to the reference contribution while leaving the correlated treatment itself unchanged.

In practice, Eq.~(\ref{eq:refcorr}) is inexpensive: no correlated calculation in the larger basis is required. Only the reference contribution must be evaluated in $L$, which is chosen to provide a near-converged reference energy, while the computationally more demanding correlation treatment remains in the small basis $S$.

%\section{Results}

We now assess this reference-corrected scheme for the non-relativistic HEAT benchmark set using the Dunning basis-set family from cc-pVDZ to aug-cc-pVQZ. Accuracy is measured in terms of mean absolute errors (MAEs) and maximum absolute errors (MaxEs) for total and atomization energies. Throughout, we compare xTC and RC-xTC against CCSD(T)-F12a or CCSD-F12a using the fixed ansatz. For the standard xTC calculations, the Jastrow factor is optimized separately in each basis set, as is standard for the method. In contrast, for RC-xTC a single Jastrow optimized in aug-cc-pVQZ is used, corresponding to the large basis $L$ in Eq.~(\ref{eq:refcorr}). Using the same Jastrow throughout the basis-set family avoids inconsistencies from basis-dependent optimization and isolates the effect of the reference correction.

We first consider the cc-pVDZ results at the level of individual systems. Figure~\ref{fig:ccsdt_scatter} compares the total- and atomization-energy errors of xTC-CCSD(T) and RC-xTC-CCSD(T). For every system in the HEAT set, the reference-corrected scheme improves the total energy. More importantly, the same trend is observed for the atomization energies, showing that the gain in total-energy accuracy translates directly into improved error cancellation between atoms and molecules. This is precisely the behavior expected from the reference/correlation decomposition above, where the dominant small-basis error was identified as a deficiency of the xTC reference contribution.

A more direct comparison with F12 is given in Fig.~\ref{fig:ccsdt_basis}, which reports the atomization-energy errors of CCSD(T)-F12 and RC-xTC-CCSD(T) for cc-pVDZ, cc-pVTZ, and cc-pVQZ. In cc-pVDZ, the spread of errors is visibly smaller for RC-xTC than for F12, indicating a more balanced description at the double-$\zeta$ level. Upon increasing the basis size, the error distributions of RC-xTC and CCSD(T)-F12 both become substantially narrower. However, in cc-pVTZ a larger fraction of systems already falls within chemical accuracy for RC-xTC, whereas CCSD(T)-F12 generally requires cc-pVQZ for a similarly large fraction of chemically accurate results. 

It can be observed that a small number of outliers remain for RC-xTC in cc-pVQZ, most notably F$_2$, O$_2$, and to a lesser extent OF. In order to shed light on these outliers, we performed RC-xTC-FCIQMC calculations in the cc-pVQZ basis on F$_2$ and F (going up to 400M walkers in the case of F$_2$), obtaining an atomization energy error of $ -0.88\; \mathrm {kcal/mol}$, i.e. a chemically accurate result.  This suggests that the failure of RC-xTC-CCSD(T) to achieve chemical accuracy for F$_2$ does not reflect a shortcoming of the reference correction itself, but rather below-par performance of the xTC-CCSD(T) method. Upon examining the FCIQMC-sampled wave function of F$_2$, we discovered that the transcorrelated molecular wave function, unexpectedly, becomes less compact than the corresponding non-TC wave function -- the coefficient of the HF determinant reduces from 0.82 to 0.69 in going from the non-TC to TC Hamiltonian. Although at present we do not understand the cause for this unexpected behaviour, we believe it could explain the relatively large error observed at the CCSD(T) level for F$_2$. 

\begin{figure*}[!t]
 \centering
 \centering
    \begin{minipage}{0.49\textwidth}
    \centering
    \includegraphics[width=\linewidth]{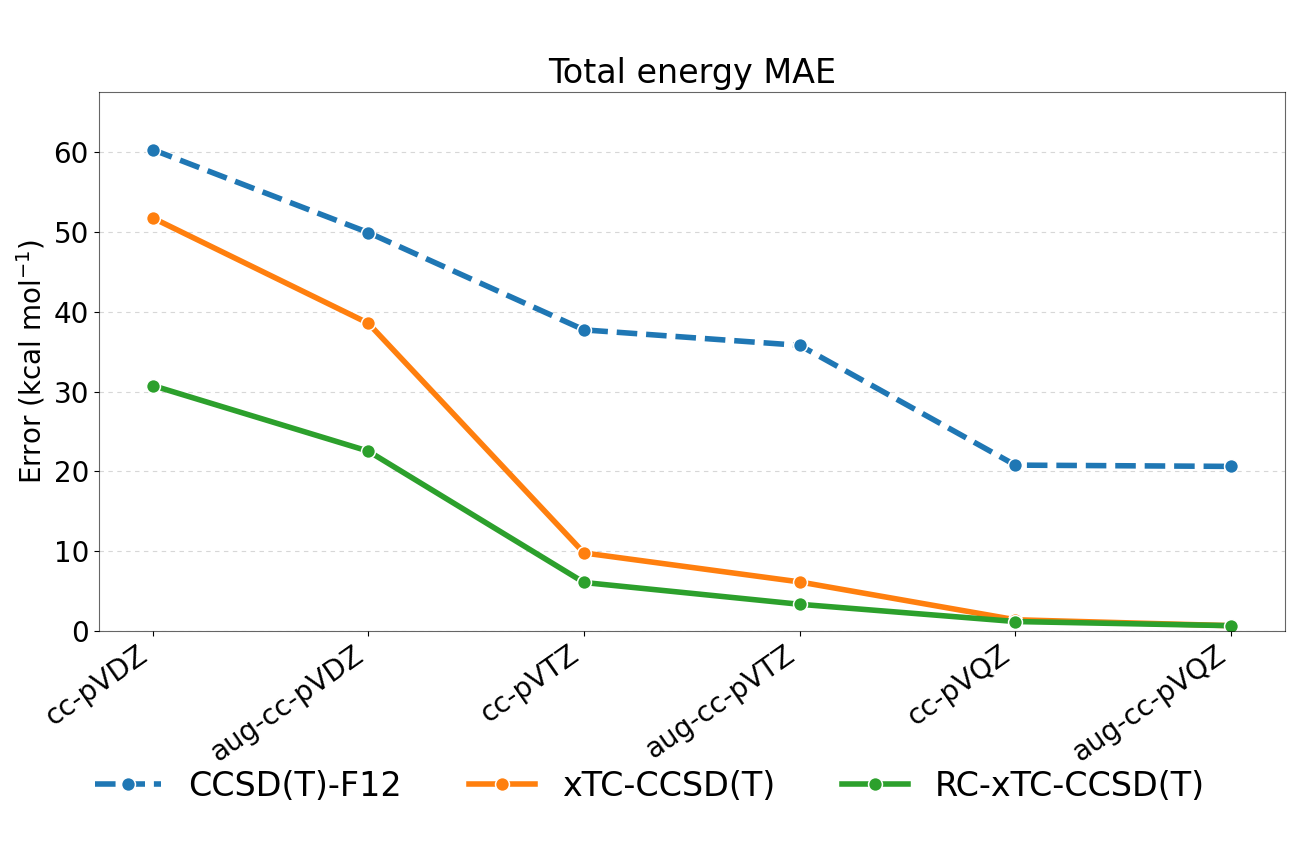}
  \end{minipage}\hfill
  \begin{minipage}{0.49\textwidth}
    \centering
    \includegraphics[width=\linewidth]{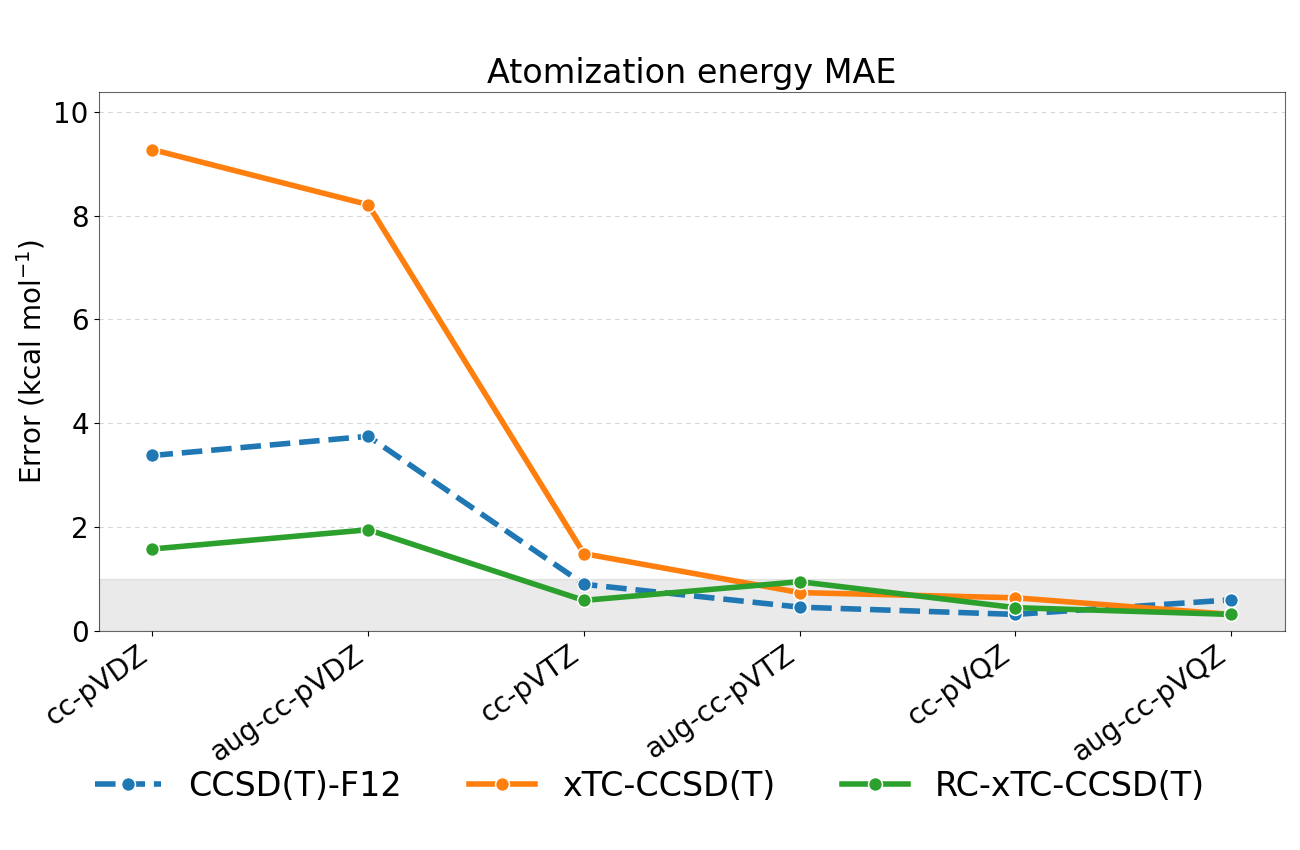}
  \end{minipage}
  \caption{
  Basis-set error statistics for the Dunning basis-set family at the CCSD(T) level. Shown are the MAEs of the total and atomization energies for CCSD(T)-F12, xTC-CCSD(T), and RC-xTC-CCSD(T), all relative to the non-relativistic HEAT reference values. While xTC already improves total energies substantially over F12, its atomization energies remain inferior in double-$\zeta$ bases. The reference correction largely removes this imbalance and yields the best atomization-energy statistics in the small-basis regime.
  }
  \label{fig:stats_ccsdt}
\end{figure*}

\begin{figure*}[!t]
 \centering
 \centering
    \begin{minipage}{0.49\textwidth}
    \centering
    \includegraphics[width=\linewidth]{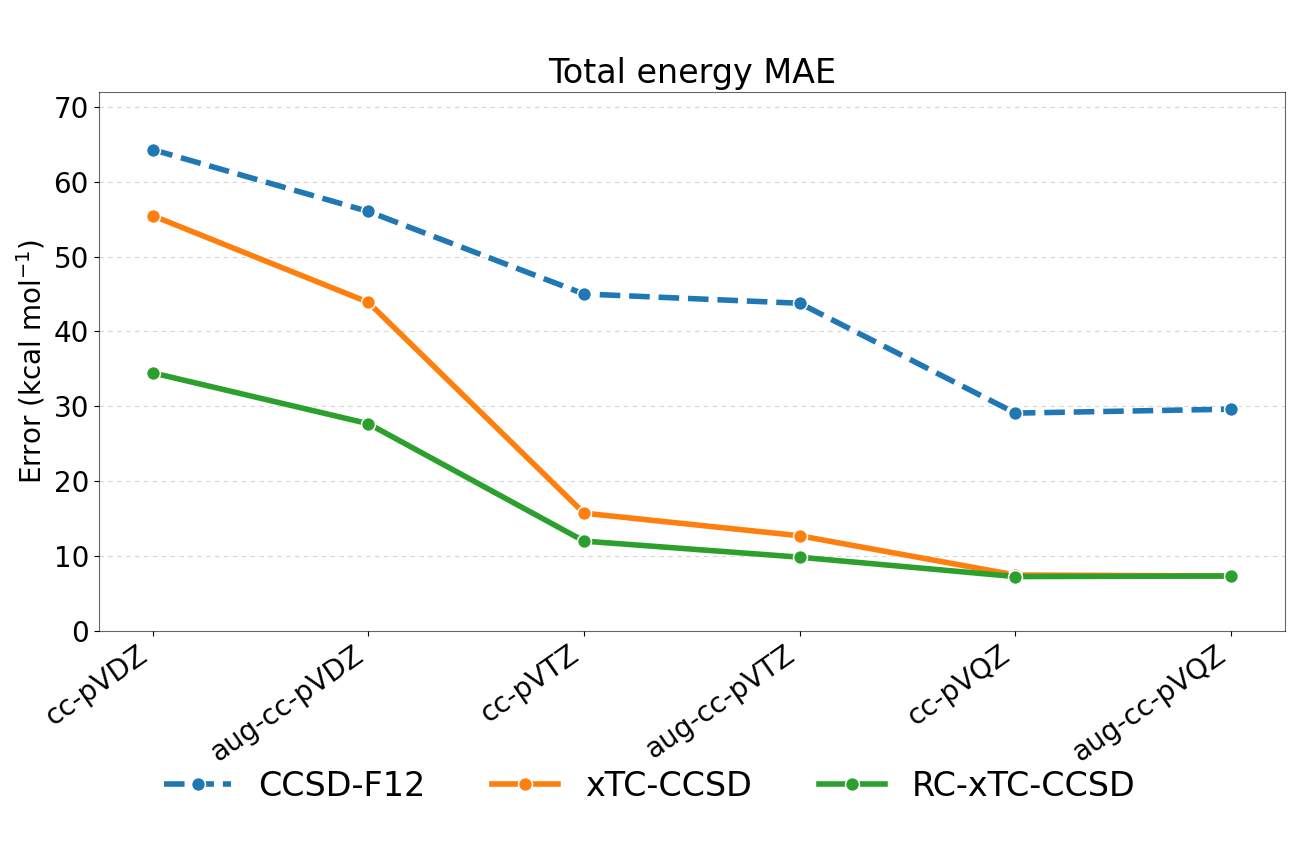}
  \end{minipage}\hfill
  \begin{minipage}{0.49\textwidth}
    \centering
    \includegraphics[width=\linewidth]{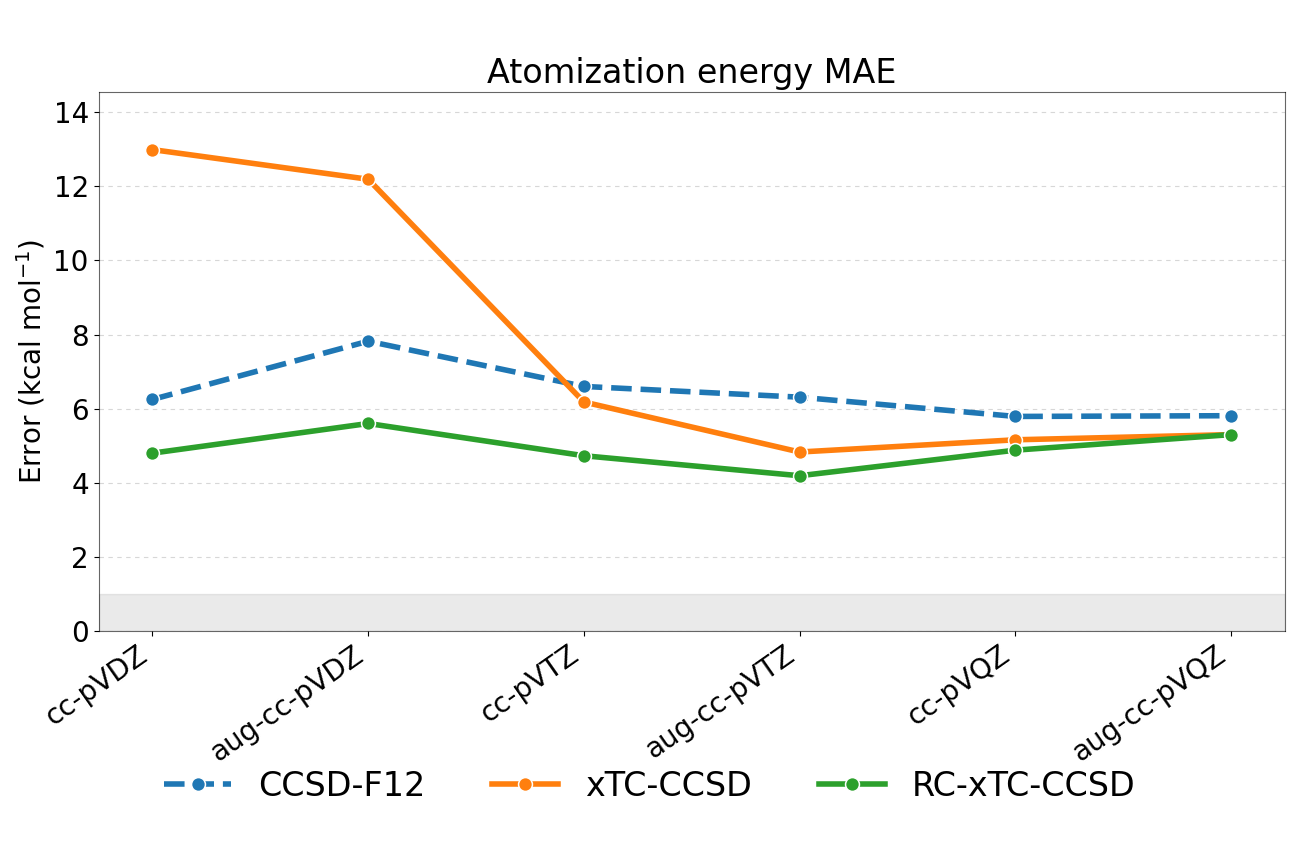}
  \end{minipage}
  \caption{
Basis-set error statistics for the Dunning basis-set family at the CCSD level. Shown are the MAEs of the total and atomization energies for CCSD-F12, xTC-CCSD, and RC-xTC-CCSD, all relative to the non-relativistic HEAT reference values. In contrast to the CCSD(T) case, neither RC-xTC nor F12 shows a pronounced basis-set convergence pattern for atomization energies, indicating that the dominant remaining error is no longer basis-set incompleteness but the intrinsic accuracy of the CCSD method itself.
  }
  \label{fig:stats_ccsd}
\end{figure*}

The basis-set dependence of the different methods is summarized quantitatively in Fig.~\ref{fig:stats_ccsdt}, with the numerical values collected in Table~\ref{tab:dunning_stats}. Several clear trends emerge. First, xTC-CCSD(T) yields significantly smaller total-energy errors than CCSD(T)-F12a across the entire Dunning family. However, this does not automatically translate into better atomization energies: in the double-$\zeta$ basis, the atomization-energy MAE and MaxE of uncorrected xTC are markedly worse than those of F12. This is exactly the imbalance anticipated from the decomposition analysis above. Second, the RC-xTC scheme improves the xTC results substantially, both for total and atomization energies. In particular, in cc-pVDZ and aug-cc-pVDZ, the corrected method not only improves upon standard xTC, but also becomes more accurate than CCSD(T)-F12 for atomization energies. Third, by construction, the difference between xTC and RC-xTC decreases with increasing basis quality, and the two become nearly indistinguishable in quadruple-$\zeta$, where the reference correction vanishes asymptotically.

\begin{table*}[!t]
\centering
\caption{
Mean absolute errors (MAEs) and maximum absolute errors (MaxEs), in kcal/mol, for total and atomization energies of the HEAT systems across the Dunning basis-set family at the CCSD(T) level. Results are shown for CCSD(T)-F12, uncorrected xTC-CCSD(T), and reference-corrected RC-xTC-CCSD(T).
}
\label{tab:dunning_stats}
\resizebox{\textwidth}{!}{%
\begin{tabular}{lcccccc|cccccc}
\hline
&
\multicolumn{6}{c|}{Total energies} &
\multicolumn{6}{c}{Atomization energies} \\
\cline{2-7}\cline{8-13}
Basis
& \multicolumn{3}{c}{MAE}
& \multicolumn{3}{c|}{MaxE}
& \multicolumn{3}{c}{MAE}
& \multicolumn{3}{c}{MaxE} \\
& F12 & xTC & RC-xTC
& F12 & xTC & RC-xTC
& F12 & xTC & RC-xTC
& F12 & xTC & RC-xTC \\
\hline
cc-pVDZ      & 60.30 & 51.75 & 30.75 & 119.67 & 105.99 & 61.43 & 3.38 & 9.27 & 1.58 & 8.94 & 14.27 & 7.38 \\
aug-cc-pVDZ  & 49.91 & 38.55 & 22.53 & 101.83 & 78.27  & 46.75 & 3.75 & 8.21 & 1.95 & 7.25 & 22.94 & 5.42 \\
cc-pVTZ      & 37.72 & 9.78  & 6.08  & 76.19  & 23.96  & 15.47 & 0.90 & 1.49 & 0.59 & 2.22 & 4.23  & 1.69 \\
aug-cc-pVTZ  & 35.78 & 6.15  & 3.35  & 69.89  & 14.62  & 8.10  & 0.46 & 0.74 & 0.95 & 1.22 & 1.81  & 1.80 \\
cc-pVQZ      & 20.78 & 1.41  & 1.19  & 40.42  & 4.32   & 3.93  & 0.32 & 0.64 & 0.45 & 1.10 & 2.65  & 1.74 \\
aug-cc-pVQZ  & 20.62 & 0.67  & 0.67  & 38.59  & 2.26   & 2.26  & 0.60 & 0.32 & 0.32 & 1.52 & 1.26  & 1.26 \\
\hline
\end{tabular}%
}
\end{table*}

\begin{table*}[!t]
\centering
\caption{
Mean absolute errors (MAEs) and maximum absolute errors (MaxEs), in kcal/mol, for total and atomization energies of the HEAT systems across the Dunning basis-set family at the CCSD level. Results are shown for CCSD-F12, uncorrected xTC-CCSD, and reference-corrected RC-xTC-CCSD.
}
\label{tab:dunning_stats_ccsd}
\resizebox{\textwidth}{!}{%
\begin{tabular}{lcccccc|cccccc}
\hline
&
\multicolumn{6}{c|}{Total energies} &
\multicolumn{6}{c}{Atomization energies} \\
\cline{2-7}\cline{8-13}
Basis
& \multicolumn{3}{c}{MAE}
& \multicolumn{3}{c|}{MaxE}
& \multicolumn{3}{c}{MAE}
& \multicolumn{3}{c}{MaxE} \\
& F12 & xTC & RC-xTC
& F12 & xTC & RC-xTC
& F12 & xTC & RC-xTC
& F12 & xTC & RC-xTC \\
\hline
cc-pVDZ      & 64.29 & 55.47 & 34.48 & 125.16 & 110.49 & 69.88 & 6.25 & 12.99 & 4.80 & 10.77 & 20.97 & 11.58 \\
aug-cc-pVDZ  & 56.03 & 43.90 & 27.69 & 110.37 & 89.37  & 58.35 & 7.82 & 12.19 & 5.60 & 14.74 & 28.43 & 12.67 \\
cc-pVTZ      & 45.01 & 15.74 & 12.01 & 91.56  & 34.28  & 27.22 & 6.60 & 6.18  & 4.73 & 12.65 & 12.96 & 11.17 \\
aug-cc-pVTZ  & 43.77 & 12.72 & 9.86  & 87.47  & 27.02  & 22.27 & 6.31 & 4.83  & 4.19 & 13.46 & 12.02 & 10.31 \\
cc-pVQZ      & 29.10 & 7.48  & 7.27  & 60.36  & 17.93  & 17.68 & 5.79 & 5.16  & 4.88 & 12.30 & 11.52 & 11.51 \\
aug-cc-pVQZ  & 29.63 & 7.35  & 7.35  & 58.88  & 16.74  & 16.74 & 5.81 & 5.30  & 5.30 & 11.94 & 11.53 & 11.53 \\
\hline
\end{tabular}%
}
\end{table*}

The corresponding comparison at the CCSD level is shown in Fig.~\ref{fig:stats_ccsd}, with the associated values given in Table~\ref{tab:dunning_stats_ccsd}. Here the picture is somewhat different. RC-xTC again lowers the total-energy errors substantially relative to uncorrected xTC and remains more accurate than CCSD-F12 throughout the basis-set sequence. For atomization energies, however, neither RC-xTC nor F12 exhibits a strong basis-set convergence pattern: the errors change comparatively little from double-$\zeta$ to quadruple-$\zeta$. The remaining discrepancy therefore appears to be dominated less by one-particle basis incompleteness than by the intrinsic accuracy of the underlying CCSD description itself. Even in this regime, RC-xTC systematically yields lower atomization-energy errors than F12. This is consistent with earlier transcorrelated studies showing that transcorrelated approaches can improve not only basis-set convergence, but also the overall accuracy relative to the corresponding conventional and explicitly correlated methods, provided that the Jastrow factor is chosen appropriately.\cite{schraivogel2021tcc,schraivogel2023tcc2,haupt2023jastrow}

Overall, these results support the central idea of the present work. In the small-basis regime, the dominant deficiency of xTC-CCSD(T) atomization energies originates from the reference contribution rather than from the correlation energy. Replacing only this component by its large-basis counterpart therefore leads to a substantial and systematic improvement, while retaining the low cost of the small-basis correlated calculation. The resulting RC-xTC scheme combines the favorable total-energy performance of xTC with a markedly improved error cancellation, making it highly competitive with standard F12 approaches and yielding lower atomization-energy errors in the double-$\zeta$ regime.

%\section{Conclusions and outlook}

In summary, we have introduced an additive reference correction for the transcorrelated method. The central idea is to retain the small-basis correlation energy, while replacing the corresponding small-basis reference contribution by its value from a larger basis,
\begin{equation}
E^{\mathrm{RC\text{-}xTC}}(\tau_L;S,L)
=
E_{\mathrm{ref}}^{\mathrm{xTC}}(\tau_L;L)
+
E_{\mathrm{corr}}^{\mathrm{xTC}}(\tau_L;S),
\end{equation}
with the larger basis $L$ chosen such that the reference energy is substantially closer to the complete-basis-set limit. In this way, the correction directly targets the dominant source of error identified in the small-basis xTC atomization energies, while preserving the low cost of the correlated calculation in the target basis $S$.

For the HEAT benchmark set, this strategy leads to a substantial improvement of total energies and, more importantly, of atomization energies in small basis sets. At the CCSD(T) level, the correction strongly improves the error cancellation relative to standard xTC and yields better atomization energies than CCSD(T)-F12a  in the double-$\zeta$ regime. At the same time, xTC and RC-xTC become progressively more similar as the basis set is enlarged, as expected from the construction of the correction.

The comparison with F12a also highlights an important qualitative difference between the methods. For CCSD(T), RC-xTC combines the favorable total-energy performance of xTC with substantially improved atomization energies in small basis sets. For CCSD, RC-xTC yields better atomization energies than F12a throughout the full Dunning basis-set sequence. This is consistent with earlier transcorrelated coupled-cluster studies showing that transcorrelated approaches can improve not only basis-set convergence, but also the overall accuracy relative to the corresponding conventional and explicitly correlated methods.\cite{schraivogel2021tcc,schraivogel2023tcc2}

At the same time, the present correction is motivated by the observed error decomposition rather than by a formal derivation. An important direction for future work is therefore the development of a more rigorous and more compact alternative that delivers a similar improvement of the reference contribution while requiring only a calculation in a single basis set. Another direction is the extension of the assessment beyond the HEAT systems. In particular, we are currently carrying out calculations for the G2 test set,\cite{curtiss1991g2,yao2020g2shci} which includes molecules containing both first-row and second-row atoms and therefore provides a broader test of the transferability of the present correction scheme. We are also investigating the applicability and performance of the current TC formalism for noncovalent interactions from the A24 data set and its later extensions.\cite{rezac2013a24,rezac2015a24}

%\section*{Supplementary material}

%The Jastrow parameter files used in the RC-xTC calculations, the RC-xTC energies and the corresponding F12 energies for all HEAT systems, basis sets, and electronic-structure methods can be found in the supplementary material.

\section*{Acknowledgments}
The authors gratefully acknowledge the support of the Max Planck Society.

%\section*{Author Declarations}

%\subsection*{Conflict of Interest}
%The authors have no conflicts to disclose.

%\subsection*{Author Contributions}
%\textbf{Johannes Hauskrecht}: Data curation (lead); Formal analysis (lead); Investigation (lead); Software (equal); Validation (lead); Visualization (equal); Writing – original draft (lead); Writing – review & editing (equal). 
%\textbf{Kristoffer Simula}: Software (equal); Writing – review & editing (equal). 
%\textbf{Yifan Cheng}: Software (equal); Writing – review & editing (equal). 
%\textbf{Evelin Martine Corvid Christlmaier}: Software (equal); Writing – review & editing (equal). 
%\textbf{Daniel Kats}: Conceptualization (equal); Project administration (equal); Resources (lead); Supervision (supporting); Visualization (equal). 
%\textbf{Ali Alavi}: Conceptualization (equal);
%Funding acquisition (equal); Project administration (equal); Supervision (lead); Visualization (equal); Writing – original draft (supporting); Writing – review & editing (equal).

%\section*{Data Availability}
%The data that support the findings of this study are available within the article and its supplementary material.

\bibliographystyle{unsrtnat}
\bibliography{bib}

\end{document}